# Intrinsic Response of Graphene Vapor Sensors

*Yaping Dan[†], Ye Lu[‡], Nicholas J. Kybert[§], A. T. Charlie Johnson[†‡]*

[†] Department of Electrical and Systems Engineering, University of Pennsylvania, Philadelphia, PA 19104, USA

[‡] Department of Physics and Astronomy, University of Pennsylvania, Philadelphia, PA 19104, USA

[§] Department of Physics, University of Warwick, Coventry, CV4 7AL, UK

Abstract

Graphene is a purely two-dimensional material that has extremely favorable chemical sensor properties. It is known, however, that conventional nanolithographic processing typically leaves a resist residue on the graphene surface, whose impact on the sensor characteristics of the system has not yet been determined. Here we show that the contamination layer both degrades the electronic properties of the graphene and masks graphene's intrinsic sensor responses. The contamination layer chemically dopes the graphene, enhances carrier scattering, and acts as an absorbent layer that concentrates analyte molecules at the graphene surface, thereby enhancing the sensor response. We demonstrate a cleaning process that verifiably removes the contamination on the device structure and allows the intrinsic chemical responses of graphene to be measured.

Graphene is a zero bandgap semimetal with extraordinary electronic[1-5] and mechanical properties[6]. Comprised of a single layer of carbon with every atom on its surface, graphene is a purely two-dimensional material and an ideal candidate for use as a chemical vapor sensor. It has been reported that the absorption of individual gas molecules onto the surface of a graphene sensor leads to a detectable change in its electrical resistance[7]. It is known, however, that typical nano-lithographic processes can leave an uncontrolled residue on graphene[8] whose impact on device transport and vapor sensing properties has not been fully explored. Moreover, the *intrinsic* sensitivity of graphene to gaseous vapors can only be determined through the use of samples where contamination from lithographic processing has been measured and verifiably removed. Graphene vapor sensors that are known to be free of chemical contamination should then be amenable to (bio)molecular surface modification to control their chemical sensitivity, as has been done for carbon nanotubes[9] and semiconductor nanowires.[10] They should also allow quantitative modeling of their sensor characteristics.[11]

Here we report on experiments where the structural and electron transport properties of a graphene field effect transistor (FET) were measured immediately after mechanical exfoliation, after contact fabrication using electron beam lithography (EBL) and thin film deposition, and after a cleaning process based on that suggested in Ref. [8]. We find that standard EBL processing left the graphene covered by a ~1-nm thick contamination layer that has a substantial impact on the transport properties and vapor sensor responses of the device. The contamination layer was removed by a high temperature cleaning process in a reducing ($H_2$/Ar) atmosphere, enabling measurements of the properties of the pristine device. Compared to the as-fabricated (contaminated) device, we find that the clean device has roughly one-third the concentration of doped carriers, four-times higher carrier mobility, and much weaker electrical response upon exposure to chemical vapors. An electrical current annealing process has been found to provide similar reductions in chemical doping and carrier scattering.[2, 12, 13]

Samples were made using mechanical exfoliation to deposit graphene sheets onto an oxidized silicon substrate (300 nm oxide thickness) with prefabricated gold alignment markers. Few-layer graphene sheets were identified by optical microscopy[14] and located with respect to the alignment markers.

Atomic Force Microscopy (AFM) was used to measure the graphene thickness. Au/Cr source and drain electrodes were then fabricated using EBL and thin film evaporation. Polymethylmethacrylate (PMMA) was used as the electron beam resist (Microchem Corp., C4 950); the resist was exposed with a 30keV electron beam at a dose of 500 μC/cm$^2$ and then developed according to manufacturer instructions. After electrode deposition by thermal evaporation and a liftoff step, the surface topography was again measured by AFM, showing evidence of contamination, presumably by residual electron beam resist. (see below). We conducted current-gate voltage (I-$V_G$) measurements of the device using the p$^{++}$ Si substrate as the back gate, and measured changes in electrical current upon exposure to chemical vapors at varying concentration. At this point the sample was cleaned by heating in flowing H$_2$/Ar (850 sccm Ar, 950 sccm H$_2$) at 400 °C for 1 h.[8] Finally, AFM, electron transport, and vapor response data were collected on the cleaned sample for comparison with that obtained from the contaminated device. For the vapor response measurements, gas flows containing analyte vapors of known concentration were created using a bubbler system, as described previously.[15] High purity nitrogen was used to flush the device between exposure to analyte-containing gas flows.

Fig.1a shows the AFM image of a typical graphene sample, shown schematically in Fig.1b. The as-exfoliated graphene film is 0.8 nm thick, and therefore presumed to be a bilayer (Fig. 1c, black line scan data). After EBL, the measured thickness is 1.8 nm (Fig. 1c, blue data), with the thickness increase attributed to the presence of PMMA residue. From the I-$V_G$ characteristic (Fig. 1d, red data), and assuming a combination of short-range and long-range carrier scattering we find a carrier mobility of 1600 cm$^2$/V-s.[1, 3] The I-$V_G$ characteristic is hysteretic, similar to that of carbon nanotube FETs, where this phenomenon was attributed to charge injection into surface traps.[16, 17] This mechanism may be relevant to graphene devices as well. The charge neutrality point (point of minimum conductivity) occurs at $V_G \approx 30\ V$, corresponding to a doped carrier density of 2.2 x 10$^{12}$/cm$^2$ at $V_G = 0$.

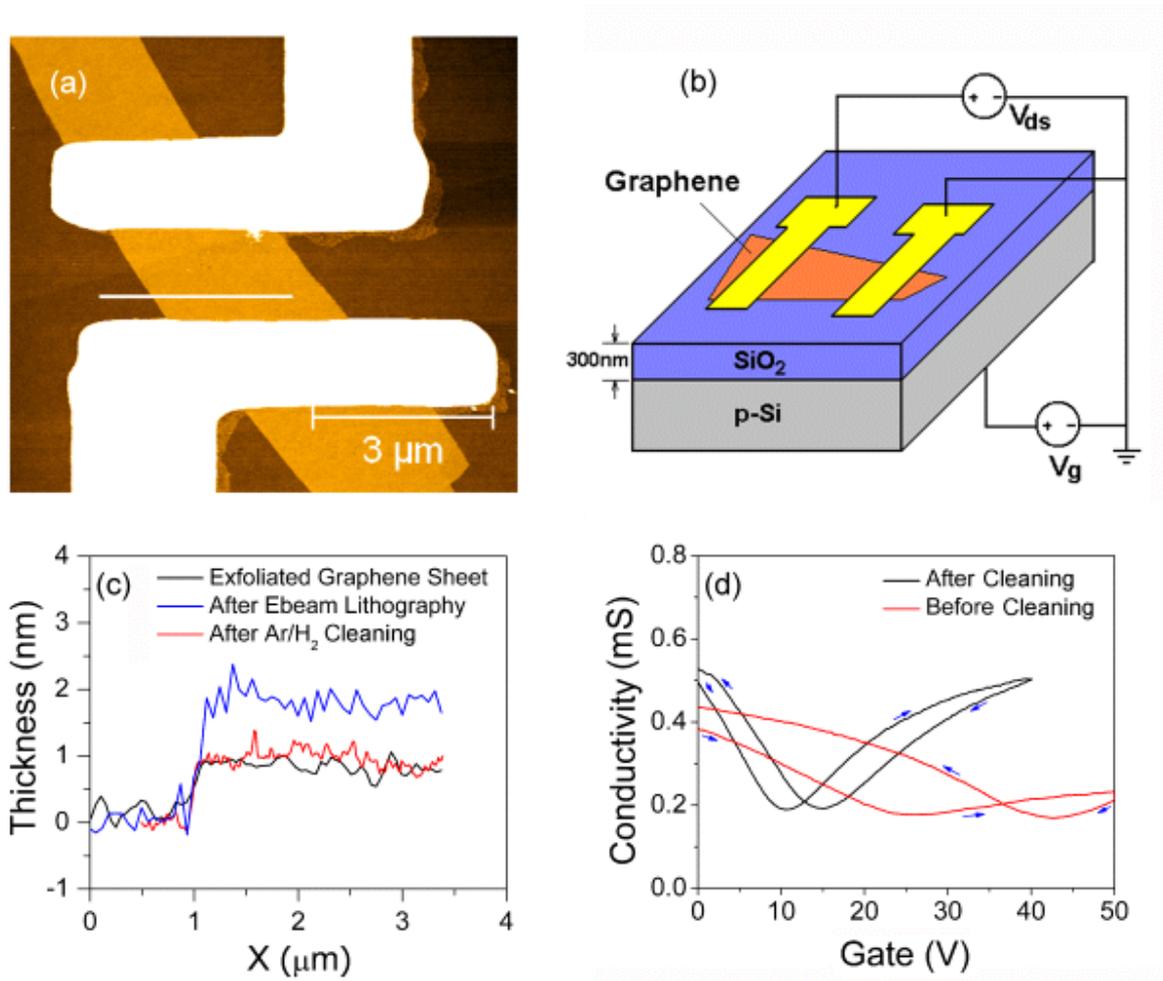

Figure 1 (a) AFM image of a graphene device. (b) Device schematic. (c) AFM line scans of the same device after exfoliation (black data; 0.8 nm thickness), after electrode fabrication by e-beam lithography (EBL) (blue data, ~ 2 nm thickness) and after a cleaning bake at 400°C in $Ar/H_2$ (red data, 0.8 nm thickness). The $Ar/H_2$ cleaning process removes the residue of the EBL resist. (d) Measured electrical conductivity versus gate voltage for the device before and after cleaning (red and black data, respectively). The cleaning step leads to significantly improved electronic properties.

We see profound changes in the AFM data and the electrical transport measurements after the cleaning bake. AFM line scans show a sample thickness of 0.8 nm, exactly equal to that of the as-exfoliated graphene (Fig 1c, red line scan data). From the I-$V_G$ data (Fig. 1d, black data), we find that the carrier mobility has increased by a factor of approximately four to 5500 $cm^2$/V-s, and that the doped carrier density at $V_G = 0$ has been reduced by two-thirds to $7.0 \times 10^{11}/cm^2$. The hysteresis in the I-$V_G$ is much smaller than that observed before the cleaning step. We conclude that the resist residue leads to carrier doping into the graphene, increased carrier scattering, and a larger density of trap states for the carrier injection that leads to greater I-$V_G$ hysteresis. The cleaning step is effective at significantly improving the structural and electronic properties of the graphene.

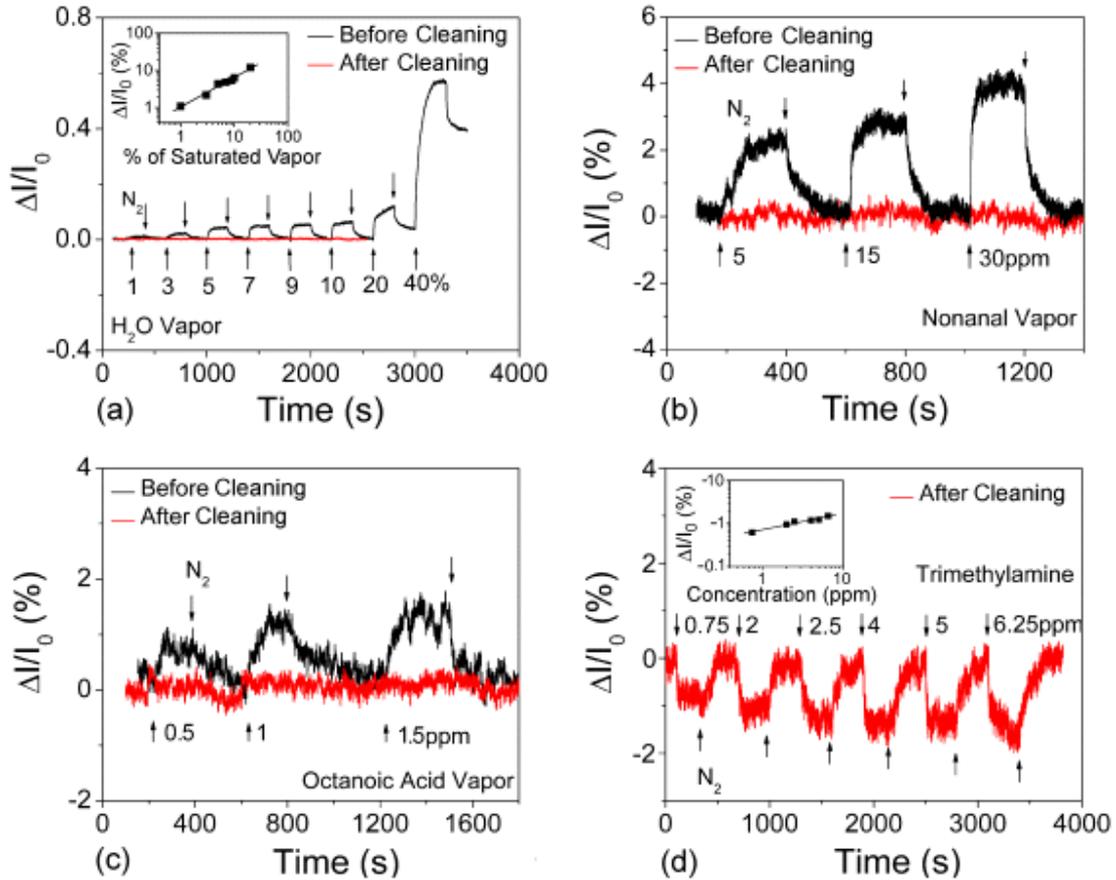

Figure 2. Measured sensor responses, before (black) and after (red) sample cleaning, to vapors of (a) water, (b) nonanal, (c) octanoic acid, and (d) trimethlyamine. The cleaning step removes resist residue from the lithography step and enables the measurement of the intrinsic responses of the graphene device. Insets: Sensor responses to water vapor and TMA show a power law dependence, with exponents of 0.4 and 0.8, respectively.

We find that the cleaning procedure leads to equally dramatic changes in the electrical response of the device upon exposure to chemical vapors at various concentrations (Fig. 2). The analytes used were water vapor, nonanal, octanoic acid, and trimethylamine (TMA). After the EBL processing and before the cleaning bake, the device shows strong electrical response to these chemical vapors, even at concentrations in the part-per-billion range in the case of octanoic acid. The responses and recovery are rapid (tens of seconds) and reversible without heating or other refreshing, although irreversible "poisoning" of the sensor response is seen upon exposure to water at a concentration of 40% of a saturated vapor (Fig. 2a). These sensor responses decrease sharply when the sample is cleaned and are thus *not* intrinsic to graphene (see below). Still, these data demonstrate that graphene vapor sensors have a number of desirable characteristics and confirm the promise of graphene for this application.

The signs of the measured vapor responses are in agreement with a model where the resist contamination acts as an unintentional "functionalization" layer that absorbs analyte molecules very near to the surface of the p-type graphene transistor, which then provides a high-sensitivity electronic readout. Water vapor is an oxidant under typical conditions (indeed, it has been suggested that the p-type behavior of graphene under ambient may be due to the effect of adsorbed water[18]), so exposure to additional water vapor is expected to increase the hole density and the current. Octanoic acid will deprotonate in the presence of adsorbed water, increasing the hole concentration (and thus the current) in the graphene by "chemical gating".[19] Trimethylamine is a proton acceptor in the presence of water, so a current decrease is expected, consistent with our observations. At this point it is unclear whether the nonanal response is consistent with this picture. The current response typically follows a power law dependence on concentration, with the exponent in the range of 0.4 – 0.8 (insets in Fig. 2). Similar power law behavior has also been reported for vapor sensors based on metal oxides[20, 21] and conducting polymer nanowires.[22]

The electrical responses to chemical vapors are reduced by one to two orders of magnitude after the cleaning bake. This observation is strong evidence that the EBL resist residue acts as an absorbent layer that concentrates molecules from the vapor within the polymer, less than 1 nm from the surface of the graphene. This behavior is not surprising since polymer films are sometimes used intentionally as analyte concentrators, for example in gravimetric vapor sensors based on surface acoustic wave devices.[23]

The results presented here illuminate a pathway towards the application of intentionally functionalized graphene devices as nanoscale sensors of molecular analytes in the vapor and liquid phase. The two-dimensional nature of graphene typically leads to devices with lower electrical noise, and thus lower detection limits, than those based on one-dimensional nanomaterials (e.g., carbon nanotubes and semiconductor nanowires).[7] For example, the data in Fig. 2 imply that graphene sensors show rapid response and recovery, and that detection of carboxylic acids and aldehydes at ppb levels should be readily attainable. The graphene surface must be clean, however, before strategies to control its chemical affinity via molecular functionalization may be exploited. Because of the similarity of the two nanomaterials, the cleaning process demonstrated here should enable the ready transfer to graphene of surface chemistry modifications previously applied to carbon nanotubes for targeted molecular sensing in the vapor and liquid phases.

This work was supported by the JSTO DTRA and the Army Research Office Grant # W911NF-06-1-0462. Support from the Nanotechnology Institute of the Commonwealth of Pennsylvania (Y.D.) and the REU program of the Laboratory for Research on the Structure of Matter (N.J.K.), NSF MRSEC DMR05-20020, are gratefully acknowledged.

References

1. Morozov, S. V. et al. Giant intrinsic carrier mobilities in graphene and its bilayer. Physical Review Letters 1 (2008).

2. Bolotin, K. I. et al. Ultrahigh electron mobility in suspended graphene. Solid State Communications 146, 351-355 (2008).

3. Chen, J. H., Jang, C., Xiao, S. D., Ishigami, M. & Fuhrer, M. S. Intrinsic and extrinsic performance limits of graphene devices on SiO2. Nature Nanotechnology 3, 206-209 (2008).

4. Geim, A. K. & Novoselov, K. S. The rise of graphene. Nature Materials 6, 183-191 (2007).

5. Novoselov, K. S. et al. Two-dimensional gas of massless Dirac fermions in graphene. Nature 438, 197-200 (2005).

6. Lee, C., Wei, X. D., Kysar, J. W. & Hone, J. Measurement of the elastic properties and intrinsic strength of monolayer graphene. Science 321, 385-388 (2008).

7. Schedin, F. et al. Detection of individual gas molecules adsorbed on graphene. Nature Materials 6, 652-655 (2007).

8. Ishigami, M., Chen, J. H., Cullen, W. G., Fuhrer, M. S. & Williams, E. D. Atomic structure of graphene on SiO2. Nano Letters 7, 1643-1648 (2007).

9. Staii, C., Chen, M., Gelperin, A. & Johnson, A. T. DNA-decorated carbon nanotubes for chemical sensing. Nano Letters 5, 1774 - 1778 (2005).

10. McAlpine, M. C. et al. Peptide-nanowire hybrid materials for selective sensing of small molecules. Journal of the American Chemical Society 130, 9583-9589 (2008).

11. Wehling, T. O. et al. Molecular doping of graphene. Nano Letters 8, 173-177 (2008).

12. Moser, J., Barreiro, A. & Bachtold, A. Current-induced cleaning of graphene. Applied Physics Letters 91 (2007).

13. Bolotin, K. I., Sikes, K. J., Hone, J., Stormer, H. L. & Kim, P. Temperature-dependent transport in suspended graphene. Physical Review Letters 101 (2008).

14. Roddaro, S., Pingue, P., Piazza, V., Pellegrini, V. & Beltram, F. The optical visibility of graphene: Interference colors of ultrathin graphite on SiO2. Nano Letters 7, 2707-2710 (2007).

15. Dan, Y. P., Cao, Y. Y., Mallouk, T. E., Johnson, A. T. & Evoy, S. Dielectrophoretically assembled polymer nanowires for gas sensing. Sensors and Actuators B-Chemical 125, 55-59 (2007).

16. Radosavljevic, M., Freitag, M., Thadani, K. V. & Johnson, A. T. Nonvolatile molecular memory elements based on ambipolar nanotube field effect transistors. Nano Letters 2, 761-764 (2002).

17. Fuhrer, M. S., Kim, B. M., Durkop, T. & Brintlinger, T. High-mobility nanotube transistor memory. Nano Letters 2, 755-759 (2002).

18. Moser, J., Verdaguer, A., Jimenez, D., Barreiro, A. & Bachtold, A. The environment of graphene probed by electrostatic force microscopy. Applied Physics Letters 92 (2008).

19. Kong, J. & Dai, H. J. Full and modulated chemical gating of individual carbon nanotubes by organic amine compounds. Journal of Physical Chemistry B 105, 2890-2893 (2001).

20. Rella, R. et al. Tin Oxide-Based Gas Sensors Prepared by the Sol-Gel Process. Sens. and Actuat. B 44, 462-467 (1997).

21. Yamazoe, N. & Shimanoe, K. Theory of Power Law for Semiconductor Gas Sensors. Sens. and Actuat. B 128, 566-573 (2008).

22. Dan, Y., Cao, Y., Mallouk, T. E., Evoy, S. & Johnson, A. T. C. Gas sensing properties of single conducting polymer nanowires and the effect of temperature. arXiv:0808.3199 (2008).

23. Wang, F. et al. Gate-variable optical transitions in graphene. Science 320, 206-209 (2008).